\documentclass[aps,showpacs,pre,superscriptaddress]{revtex4}%
\usepackage{amsfonts}
\usepackage{amsmath}
\usepackage{mathrsfs}
\usepackage{amssymb}
\usepackage{wasysym}
\usepackage{subfigure}
\usepackage{graphicx}%
\providecommand{\U}[1]{\protect\rule{.1in}{.1in}}

\begin{document}
\title{An analytical approach to quantum phase transitions of ultracold Bose systems in bipartite optical lattices: Along the avenue of Green's function }
\author{Zhi Lin}
\affiliation{Department of Physics, Shanghai University, Shanghai 200444, P.R. China}
\affiliation{Department of Physics, State Key Laboratory of Surface Physics and Laboratory of Advanced Materials, Fudan University, Shanghai 200433, P.R. China}
\author{Jun Zhang}
\affiliation{Department of Physics, Shanghai University, Shanghai 200444, P.R. China}
\author{Yan Chen}
\affiliation{Department of Physics, State Key Laboratory of Surface Physics and Laboratory of Advanced Materials, Fudan University, Shanghai 200433, P.R. China}
\author{Ying Jiang}
\thanks{Corresponding author}
\email{yjiang@shu.edu.cn}
\affiliation{Department of Physics, Shanghai University, Shanghai 200444, P.R. China}
\affiliation{Key Lab for Astrophysics, Shanghai 200234, P.R. China}

\begin{abstract}
In this paper, we present a generalized Green's function method which can be used to investigate the quantum phase transitions analytically in a systematic way for ultracold Bose systems in bipartite optical lattices. As an example, to the lowest order, we calculate the quantum phase boundaries of the localized states (Mott insulator or charge density wave) of an ultracold Bose system in a $d$-dimensional hypercubic optical lattice with nearest-neighbor repulsive interactions. Due to the inhomogeneity of the system, in the generalized Green's function method, cumuants on different sublattices are calculated separately, together with re-summed Green's function technique, the analytical expression of the phase boundaries of the localized phases in the system is presented.

\end{abstract}


\maketitle

\section{Introduction}

Without any doubt, the physics of ultracold Bose systems in optical lattices has become one of the hottest research fields for more than a decade \cite{greiner} due to their novelty and various potential applications as so-called quantum simulations \cite{Immanuel-Bloch,lewenstein}. Among the research topics on ultracold Bose systems in optical lattices, to determine the ground states and the corresponding quantum phase transitions \cite{sachdev-qptbook} is one of the major and important problems \cite{fisher,zoller,greiner}.

As is known \cite{zoller}, an ultracold spinless Bose system in a homogeneous optical lattice can be described by the elegant Bose-Hubbard Hamiltonian \cite{Immanuel-Bloch,fisher}
\begin{equation}
\hat{H}_{\rm BH}=\hat{H}_{1}+\hat{H}_{0} \label{bose-hubbard-hamiltonian}
\end{equation}
with the diagonal part $\hat{H}_{0}$ and the hopping part
$\hat{H}_{1}$ being
\begin{equation}
\hat{H}_{0}=\sum_{i}\frac{U}{2}\hat{n}_{i}(\hat{n}_{i}-1)-\mu\hat{n}_{i},\,\,\,\hat{H}_{1}=-J\sum_{\langle
i,j\rangle}\hat{a}_{i}^{\dagger}\hat{a}_{j}, \label{equation2}
\end{equation}
respectively, where $\hat{n}_{i}=
\hat{a}_{i}^{\dagger}\hat{a}_{i}$\ is the particle number operator
on site $i$. $J$ is the hopping amplitude for the bosons between
the nearest-neighbor sites $i$ and $j$ while $U$ denotes the
on-site repulsion and $\mu$ is the chemical potential. To investigate this simple yet nontrivial Hamiltonian analytically, mean field theory \cite{fisher} and strong-coupling expansion \cite{freericks1} can be employed. However, comparison with the Monte Carlo simulation results \cite{capogrosso1} shows that the former is underestimates the phase boundary while the latter goes in the opposite direction.

By treating the hopping parameter as a perturbation, based on the effective potential and Rayleigh-Schr\"odinger perturbation theory, an alternative analytical approach has been developed \cite{axel-09}. With this novel systematic approach, the superfluid-Mott-insulator quantum phase boundaries for ultracold Bose systems in square and cubic lattices \cite{axel-09} as well as in non-rectangular optical lattices (including triangular, hexagonal, and kagom\'e lattices) \cite{jiang-pra-1} have been determined analytically, in good agreement with numerical results \cite{Nikalas Teichmann1}. Based on the same perturbation treatment, by utilizing the cumulants expansion and re-summed Green's function technique \cite{Metzner,ohliger}, the time-of-flight absorption pictures and the corresponding visibility for ultra-cold bosons in a triangular optical lattice have been calculated analytically \cite{jiang-pra-2}, the comparison between our analytical results and the experimental data \cite{becker1} exhibits a qualitative agreement.

Along with the progress of experimental methods and the deepening of theoretical studies, in recent year, more and more efforts have been devoted to more complex systems, including systems with long-range interactions \cite{lahaye,lauer1,trefzger,naini,schauss1}, multi components \cite{altman1,soltan-panahi1}, frustrations \cite{eckardt-2010,pielawa,ye-2012}, or superlattice structures \cite{piel1,sebby-strabley1,folling1,cheinet1,jo1}. These systems, in pratice, should be treated as systems in inhomogeneous lattices, i.e. lattices consisting of multi sublattices.
However, the above mentioned hopping parameter perturbation approaches so far can only be used to cope with  systems in homogeneous lattices, and need to be generalized.

Last year, by taking Bose systems in square and cubic superlattices as examples, Wang {\it et al.} \cite{wang1} have developed a generalized effective potential field theory which can be used to compute the quantum phase diagrams for multi-component Bose systems. In this paper, based on the same perturbation philosophy, we are going to approach  bipartite sublattice problems along an alternative route via the methods of Green's functions. As is known \cite{jiang-pra-2}, the Green's function can directly decrypt the information of the time-of-flight absorption picture and the visibility of a Bose system in an optical lattice. Meanwhile, for second order phase transitions, the divergence of Green's functions may also be used to locate the phase boundaries \cite{Kleinert}. In order to illustrate the generalized Green's function method, we choose a scalar Bose system in a $d$-dimensional hypercubic lattice with nearest-neighbor interaction as a toy model.

The remainder of the paper is organized as follows: In order to make the paper more self-contained, we would like first to give a brief review of the cumulants expansion and re-summed methods of Green's function in Section II. Section III is devoted to detail discussion for the localized states of the nearest-neighbor interacting Bose system in a $d$-dimensional hypercubic lattice. In Section IV, the effort is mainly concentrated on generalizing the Green's function methods for multi-component systems. As an example, we present the phase boundaries of the localized ground states discussed in the Section III via the extended Green's function technique. Discussion is in the Section V.

\section{A brief review of the cumulant expansion and re-summed Green's function methods}

A scalar Bose system trapped in a homogeneous optical lattice is depicted by the Bose-Hubbard Hamiltonian in Eq.(\ref{bose-hubbard-hamiltonian}),
the corresponding one particle Green's function $G(\tau^{\prime},j^{\prime}\mid \tau,j)$ of this system is defined as
\begin{equation}
G(\tau^{\prime},j^{\prime}\mid \tau,j)=\,\langle \hat{T}_{\tau}
[\hat{a}^{\dagger}_{j^{\prime}}(\tau^{\prime}) \hat{a}_j(\tau)]
\rangle,
\end{equation}
where $\tau$ is imaginary time and $\hat{T}_{\tau}$ is the
imaginary  time ordering operator, $\langle\cdot\rangle$ denotes the average with respect to the full Hamiltonian in Eq.(\ref{bose-hubbard-hamiltonian}).

In fact, the information of the momentum space distribution function $n(\mathbf k)$ of the system, which can be revealed by experiment directly via time-of-flight technique \cite{kashurnikov1}, is encrypted in the one particle Green's function. The momentum space distribution function reads
\begin{equation}
n(\mathbf k)=N_S\mid\emph{w}(\mathbf k)\mid^2\langle
\hat{a}^{\dagger}_{\mathbf k} \hat{a}_{\mathbf k} \rangle,
\label{equation7}
\end{equation}
where $\hat{a}^{\dagger}_{\mathbf k}$ and $\hat{a}_{\mathbf k}$ are the Fourier transformation of the operators in Eq.(\ref{bose-hubbard-hamiltonian}), i.e.
\begin{equation}
\hat{a}^{\dagger}_i=\frac {1}{\sqrt{N_S}}\sum_{\mathbf
k_{1}}\hat{a}^{\dagger}_{\mathbf k_{1}}e^{-i\mathbf k_{1}
\cdot\mathbf r_{i}},\,\,\,\hat{a}_j=\frac
{1}{\sqrt{N_S}}\sum_{\mathbf k_{2}}\hat{a}_{\mathbf
k_{2}}e^{i\mathbf k_{2}\cdot\mathbf r_{j}}, \label{equation6}
\end{equation}
$N_S$ is the total of the lattice number, and $w({\mathbf k})$ is the corresponding Wannier function in momentum space.
It is observed that $\langle \hat{a}^{\dagger}_{\mathbf k}
\hat{a}_{\mathbf k} \rangle = \lim_{\tau\downarrow 0}G(\tau
\mid0,\mathbf k)$, while $G(\tau \mid 0, \mathbf k)=\,\langle
\hat{T}_{\tau}[\hat{a}^{\dagger}(\tau)_{\mathbf k}
\hat{a}(0)_{\mathbf k}] \rangle$ is exactly the one particle Green's function in the momentum
space which is the Fourier transformation of $G(\tau^{\prime},j^{\prime}\mid \tau,j)$. Moreover, according to Laudau's theory of phase transitions \cite{Kleinert,Quantum-Field-Theory}, for second order phase transitions, which is the case for SF-MI phase transition, the divergence of Green's function reveals the location of phase boundaries.
Hence, it is necessary to calculate the one particle Green's function explicitly.

However, due to the non-commutativity of the hopping part and the local part of the
Hamiltonian (\ref{bose-hubbard-hamiltonian}), the Green's functions may be calculated perturbatively via the
Dirac representation by treating the hopping term in the Hamiltonian
(\ref{bose-hubbard-hamiltonian}) as the perturbation part \cite{Metzner,ohliger,jiang-pra-2}, for the opposite limit
will not lead to a MI-SF phase transition at all \cite{stoof}. In Dirac picture, the one particle Green's function reads
\begin{eqnarray}
G(\tau^{\prime},j^{\prime}\mid
\tau,j)&=&\,\frac{\text{Tr}\{{{e^{-\beta
\hat{H}_{0}}}\hat{T}_{\tau}[\hat{a}^{\dagger}_{j^{\prime}}(\tau^{\prime})\hat{a}_{j}(\tau)\hat{u}(\beta,0)]}\}}
{\{\text{Tr}{e^{-\beta \hat{H}_{0}}}\hat{u}(\beta,0)\}},
\label{equation40}
\end{eqnarray}
where
$\hat{O}(\tau)=e^{\tau\hat{H}_{0}}\hat{O}e^{- \tau\hat{H}_{0}}$,
and $\hat{u}(\beta,0)=\hat{T}_{\tau}\left[\exp\left(\int^{\beta}_{0}d\tau\sum_{\langle
i,j \rangle} J
\hat{a}^{\dagger}_{i}(\tau)\hat{a}_{j}(\tau)\right)\right]
$ is the evolution operator (setting
$\hbar=1$) \cite{M. Peskin}.

The perturbative expansion of $\hat{u}(\beta,0)$ results in terms like
\begin{eqnarray}
\frac{1}{n!}\sum_{i_{1},j_{1},\cdots,i_{n},j_{n}}J_{i_{1},j_{1}}\cdots
J_{i_{n},j_{n}}\int^{\beta}_{0}d\tau_{1} \cdots
\int^{\beta}_{0}d\tau_{n}
\langle\hat{T}_{\tau}[\hat{a}^{\dagger}_{j^{\prime}}(\tau^{\prime})\hat{a}_{j}(\tau)\hat{a}^{\dagger}_{i_{1}}(\tau_{1})
\hat{a}_{j_{1}}(\tau_{1})\cdots\hat{a}^{\dagger}_{i_{n}}(\tau_{n})\hat{a}_{j_{n}}(\tau_{n})]\rangle_{0}
\end{eqnarray}
in the expanded expression of $G(\tau^{\prime},j^{\prime}\mid \tau,j)$, here
$\langle \hat{O}\rangle_{0}$ denotes an average quantity with
respect to the unperturbed part $\hat{H}_0$ of the Hamiltonian, $J_{ij}$ reads
\begin{equation}J_{ij}=\begin{cases}
 J, \;\;\;& \text{if {\it i, j} are nearest-neighbors of each other,}\\
 0, \;\;\;&\text{otherwise.}
\end{cases}
\end{equation}
Then the problem of calculating one particle Green's function $G(\tau^{\prime},j^{\prime}\mid \tau,j)$ is transformed to a problem of calculating the $n$-particle Green's function with respect to $\hat{H}_0$
\begin{equation}\label{equation11}
G^{(0)}_{n}(\tau^{\prime}_{1},i^{\prime}_{1};\cdots\tau^{\prime}_{n},i^{\prime}_{n}\mid
\tau_{1},i_{1};\cdots\tau_{n},i_{n}) =\langle
\hat{T}_{\tau}[\hat{a}^{\dagger}_{i^{\prime}_{1}}(\tau^{\prime}_{1})\hat{a}_{i_{1}}(\tau_{1})\cdots
\hat{a}^{\dagger}_{i^{\prime}_{n}}(\tau^{\prime}_{n})\hat{a}_{i_{n}}(\tau_{n})]\rangle_{0},
\end{equation}
However, the Wick's theorem becomes invalid due to the locality of the non-perturbed
part of the Hamiltonian (\ref{bose-hubbard-hamiltonian}). Instead, we use the theory of linked-cluster
expansion \cite{Metzner,ohliger,jiang-pra-2} to expand the $n$-particle
Green's function in terms of the cumulants
\begin{equation}
C^{(0)}_m(\tau^{\prime}_{1},\cdots,\tau^{\prime}_{m}\mid\tau_{1},\cdots\tau_{m})
=\langle
\hat{T}_{\tau}[\hat{a}^{\dagger}(\tau^{\prime}_{1})\hat{a}(\tau_{1})\cdots
\hat{a}^{\dagger}(\tau^{\prime}_{m})\hat{a}(\tau_{m})]\rangle_{0}
\label{cumulants}
\end{equation}
in which the particle operators are all at the same site, since each annihilation operator in $G^{(0)}_{n}(\tau^{\prime}_{1},i^{\prime}_{1};\cdots\tau^{\prime}_{n},i^{\prime}_{n}\mid
\tau_{1},i_{1};\cdots\tau_{n},i_{n})$ must be paired by a creation operator at the same site due to the locality of the eigenstates of $\hat{H}_0$.

The decomposition of the above mentioned
$n$-particle Green's function in terms of these cumulants is quite straightforward, for instance, the decompositions of
one and two particle Green's functions are shown in the following
\begin{equation}\label{equation14}
G^{(0)}_{1}(\tau^{\prime},i^{\prime}\mid\tau,i)=\delta_{i^{\prime},i}\,C^{(0)}_{1}(\tau^{\prime}\mid
\tau)
\end{equation}
\begin{eqnarray}\label{equation15}
G^{(0)}_{2}(\tau^{\prime}_{1},i^{\prime}_{1};\tau^{\prime}_{2},i^{\prime}_{2}\mid
\tau_{1},i_{1};\tau_{1},i_{1})&=&
\delta_{i^{\prime}_{1},i_{1}}\delta_{i^{\prime}_{1},i^{\prime}_{2}}\delta_{i_{1},i_{2}}\,C^{(0)}_{2}(\tau^{\prime}_{1},\tau^{\prime}_{2}\mid \tau_{1},\tau_{2})\nonumber \\
& &+\delta_{i^{\prime}_{1},i_{1}}\delta_{i^{\prime}_{2},i_{2}}C^{(0)}_{1}(\tau^{\prime}_{1}\mid \tau_{1})C^{(0)}_{1}(\tau^{\prime}_{2}\mid \tau_{2})\nonumber \\
& &+\delta_{i^{\prime}_{1},i_{2}}\delta_{i^{\prime}_{2},i_{1}}C^{(0)}_{1}(\tau^{\prime}_{1}\mid \tau_{2})C^{(0)}_{1}(\tau^{\prime}_{2}\mid \tau_{1}).
\end{eqnarray}
And all these expressions can be diagrammatized as
\begin{equation}\label{equation16}
C^{(0)}_{1}(\tau^{\prime}\mid
\tau)=\raisebox{-0.4cm}{\includegraphics[width=2cm]{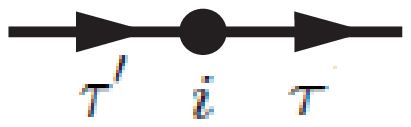}},
\end{equation}
\begin{equation}\label{equation17-old}
C^{(0)}_{2}(\tau^{\prime}_{1},\tau^{\prime}_{2}\mid
\tau_{1},\tau_{2})=\raisebox{-0.7cm}{\includegraphics[width=2cm]{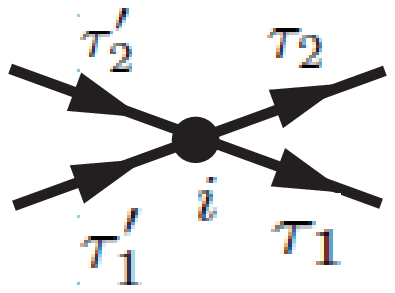}},
\end{equation}
and the hopping parameter $J_{ij}$ may be represented correspondingly as
\begin{equation}\label{equation16-1}
J_{ij}=\raisebox{-0.3cm}{\includegraphics[width=2cm]{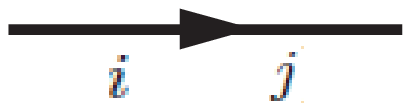}}.
\end{equation}
When considering the cancellation effect
of the denominator in Eq.(\ref{equation40}), with the help of the cumulants and the diagram rules mentioned above, the one particle
Green's function $G(\tau^{\prime},j^{\prime}\mid \tau,j)$ only
consists of connected diagrams.

Perturbative calculation of Green's function needs us to re-group all the connected diagrams based on the property of loops in diagrams instead of the order of $J/U$, since finite-order perturbative calculation always presents a polynomial and cannot reflect the divergence of Green's function at phase transition point. This is so-called re-summed Green's function technique \cite{ohliger}. By the use of the cumulants expansion and the re-summed Green's function method, we have calculated, to the lowest order, the time-of-flight absorption pictures and the corresponding visibility for ultra-cold Bose system in triangular lattice optical lattice analytically \cite{jiang-pra-2}, our results exhibit a qualitative agreement with experimental data \cite{becker1}.

\section{The localized states of Bose systems with nearest-neighbor interaction in hypercubic lattices}

However, the above reviewed Green's function method treats all lattice sites equally, and thus can only deal with systems with homogeneous lattice structures. In order to investigate systems with multi-sublattice structures, the Green's function technique needs to be generalized.

\begin{figure}[h!]
\centering
\includegraphics[width=5cm]{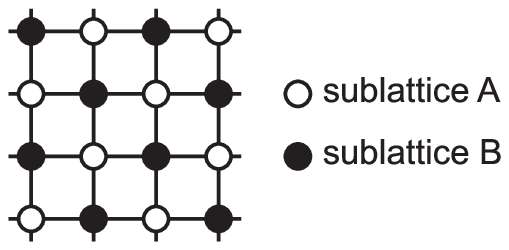}
\caption{Sketch of the sublattice structure of the nearest-neighbor interacting Bose system in square lattice.}
\label{square-lattice}
\end{figure}

In order to illustrate the generalization more clearly, we concentrate on systems of ultra-cold bosons with nearest-neighbor repulsion in $d$-dimensional hypercubic optical lattice, these systems are described by the extended Bose Hubbard Hamiltonian
\begin{equation}
\hat{H}_{\rm EBHM}=\hat{H}_{1}+\hat{H}_{0} \label{equation1},
\end{equation}
where
\begin{equation}
\hat{H}_{0}=\sum_{i}\frac{U}{2}\hat{n}_{i}(\hat{n}_{i}-1)-\mu\hat{n}_{i}+V\sum_{\langle
i,j\rangle}\hat{n}_{i}\hat{n}_{j},\,\,\,\hat{H}_{1}=-J\sum_{\langle
i,j\rangle}\hat{a}_{i}^{\dagger}\hat{a}_{j}. \label{twopartshamiltonian}
\end{equation}
Now, the local part $\hat{H}_0$ includes also the nearest-neighbor repulsive interaction $V$. Due to this nearest-neighbor interaction, the hypercubic lattice is naturally divided into two sublattices, i.e. it has bipartite lattice structure, as sketched in Fig.(\ref{square-lattice}).

The subtle balance between the hopping parameter $J$, onsite repulsion $U$ and nearest-neighbor repulsion $V$ introduces rich phase structures to this system. Again, we treat the hopping parameter $J$ as a perturbation. However, even if $J$ takes the value of 0, the competition between $U$ and $V$ will lead to different localized phases at zero temperature. In order to develop the generalized Green's function method and further calculate the phase boundaries of the localized phases as well, the localized phases at $J=0$ need to be clarified first. This can be achieved easily by the following analysis.

Let us play a game of throwing bosons into the lattice. Suppose the total number of lattice sites is $N$. Due to the nearest-neighbor repulsion, the first $\frac{N}{2}$ particles will fill up either sublattice A or sublattice B, let us say, sublattice A. Where the next $\frac{N}{2}$ bosons go is determined by the relative strength of $U$ and $zV$ ($z$ is the number of nearest-neighbor sites, in present case $z=2d$).

Let us first assume $U>zV$. Then, the second group of $\frac{N}{2}$ bosons will fill up the sublattice B. Without loss of generality, let the filling factor $n_{A}$ of sublattice A be always larger than or equal to $n_{B}$ of sublattice B. Apparently, in this case ,the third group will go to sublattice A, then followed by filling up sublattice B, and so on so forth. In other words, $n_{A}-n_{B}$ is always 1 or 0. We denote these localized states by $|n_A,n_B\rangle$, in which $|n,n\rangle$ is called Mott-insulator states, while $|n+1,n\rangle$ is the so-called charge-density-wave (CDW) states. All these localized states are eigenstates of $\hat{H}_0$, the corresponding eigenenergies read
\begin{eqnarray}
&&\hat{H}_{0}\left|n_{A},n_{B}\right\rangle= \frac{N}{2}E_{n_{A},n_{B}}\left|n_{A},n_{B}\right\rangle, \nonumber \\
&&E_{n_{A},n_{B}}=\frac{U}{2}n_{A}(n_{A}-1)+\frac{U}{2}n_{B}(n_{B}-1)+zVn_{A}n_{B}-\mu(n_{A}+n_{B}).
\label{eigenvalue}
\end{eqnarray}
The constraint on the chemical potential $\mu$ in each Mott region is determined by solving the equations $E_{n,n}\leq E_{n+1,n}$ and $E_{n,n}\leq E_{n,n-1}$, this leads to
\begin{equation}
U(n-1)+zVn\leq \mu \leq Un + zVn.
\end{equation}
Similarly, we get for the CDW regions that
\begin{equation}
U(n-1)+zV(n-1)\leq \mu \leq U(n-1) + zVn.
\end{equation}

For the case of $U<zV$, it is not difficult to figure out that bosons always go to sublattice A, i.e. $n_A=n$ while $n_B=0$, the ground state is always CDW state $|n,0\rangle$, and the chemical potential for each filling factor $n$ satisfies
\begin{equation}
U(n-1)\leq \mu \leq Un.
\end{equation}

When $U=zV$, it is known \cite{iskin-2011} that the $(n+1,n)$ CDW state becomes degenerate with the $(2n+1,0)$ CDW state, and the $(n,n)$ Mott state becomes degenerate with the $(2n,0)$ CDW state. This case will not be discussed in the present paper.

\section{The generalized Green's function method}

With the ground-state configurations of $\hat{H}_0$ in hand, we can now switch on the hopping parameter $J$ gradually. The localized ground states will be softened by the quantum fluctuations and will transit into some delocalized and compressible phases (either superfluid phase or supersolid phase) when $J$ goes beyond some critical value. These phase transitions are entirely caused by the quantum fluctuations due to the noncommutativity of the local and hopping parts in the Full Hamiltonian in Eq. (\ref{equation1}), hence are the so-called quantum phase transitions \cite{sachdev-qptbook}. As is confirmed \cite{Kovrizhin}, these quantum phase transitions are second-order phase transitions. The actual phase boundaries of the localized phases need to be determined quantitatively in an analytical way, and we are going to tackle this problem along the avenue of Green's function. This also provides us an opportunity to illustrate the generalized re-summed Green's function method clearly.

As is known \cite{Metzner,ohliger,jiang-pra-2}, the Green's function method reviewed in the preceding section is for homogeneous lattice structures, hence needs to be generalized in order to treat the present problem.

We again look the hopping parameter upon as perturbation, i.e. the unperturbed states are the localized phases discussed in the preceding section. Formally, the one-particle Green's function $G^{\rm inh}(\tau',j'|\tau,j)$ for the extended Bose-Hubbard model takes the same form as in Eq.(\ref{equation40}) for the case of homogeneous lattice, but with an extra term of nearest-neighbor repulsive interaction in $\hat{H}_0$ expressed in Eq.(\ref{twopartshamiltonian}). In the same manner, the evolution operators $\hat{u}(\beta,0)$ in the numerator and denominator should also be expanded perturbatively with respect to the hopping amplitude $J$, this again transforms the problem of calculating the one-particle Green's function $G^{\rm inh}(\tau',j'|\tau,j)$ to the problem of calculating $n$-particle Green's functions $G^{(0)}_{n}(\tau^{\prime}_{1},i^{\prime}_{1};\cdots\tau^{\prime}_{n},i^{\prime}_{n}\mid
\tau_{1},i_{1};\cdots\tau_{n},i_{n}) =\langle
\hat{T}_{\tau}[\hat{a}^{\dagger}_{i^{\prime}_{1}}(\tau^{\prime}_{1})\hat{a}_{i_{1}}(\tau_{1})\cdots
\hat{a}^{\dagger}_{i^{\prime}_{n}}(\tau^{\prime}_{n})\hat{a}_{i_{n}}(\tau_{n})]\rangle_{0}$ with respect to $\hat{H}_0$ in Eq.(\ref{twopartshamiltonian}). Due to the locality of the states determined by $\hat{H}_0$, $G^{(0)}_{n}(\tau^{\prime}_{1},i^{\prime}_{1};\cdots\tau^{\prime}_{n},i^{\prime}_{n}\mid
\tau_{1},i_{1};\cdots\tau_{n},i_{n})$ may be calculated analytically via cumulants expansion technique.

But, due to the nearest-neighbor interactions $V$, sublattices $A$ and $B$ are distinguishable, and the local properties of lattice sites belonging to different sublattices are different, hence, in calculating the $n$-particle Green's functions $G^{(0)}_{n}(\tau^{\prime}_{1},i^{\prime}_{1};\cdots\tau^{\prime}_{n},i^{\prime}_{n}\mid
\tau_{1},i_{1};\cdots\tau_{n},i_{n})$, cumulants on different sublattices need to be discriminated. For the present cases, there are two different groups of cumulants
\begin{equation}
C^{(0)}_{nA}=\langle
\hat{T}_{\tau}[\hat{a}^{\dagger}_A(\tau^{\prime}_{1})\hat{a}_A(\tau_{1})\cdots
\hat{a}^{\dagger}_A(\tau^{\prime}_{})\hat{a}_A(\tau_{})]\rangle_{0}
\end{equation}
and
\begin{equation}
C^{(0)}_{nB}=\langle
\hat{T}_{\tau}[\hat{a}^{\dagger}_B(\tau^{\prime}_{1})\hat{a}_B(\tau_{1})\cdots
\hat{a}^{\dagger}_B(\tau^{\prime}_{})\hat{a}_B(\tau_{})]\rangle_{0}
\end{equation}
in which $a_{A}$ ($a^{\dagger}_A$) and $a_B$ ($a^{\dagger}_B$) are operators on sublattices $A$ and $B$, respectively. They may be diagrammatized, for instance, as follows
\begin{equation}
C^{(0)}_{1A}(\tau^{\prime}\mid
\tau)=\raisebox{-0.4cm}{\includegraphics[width=2cm]{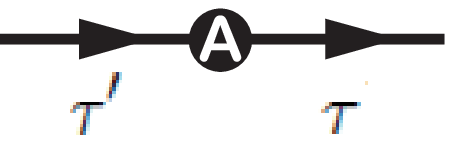}},\;\;\;\;\;C^{(0)}_{1B}(\tau^{\prime}\mid
\tau)=\raisebox{-0.4cm}{\includegraphics[width=2cm]{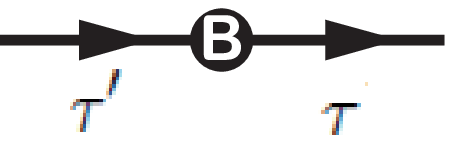}},
\end{equation}
\begin{equation}
C^{(0)}_{2A}(\tau^{\prime}_{1},\tau^{\prime}_{2}\mid
\tau_{1},\tau_{2})=\raisebox{-0.7cm}{\includegraphics[width=1.8cm]{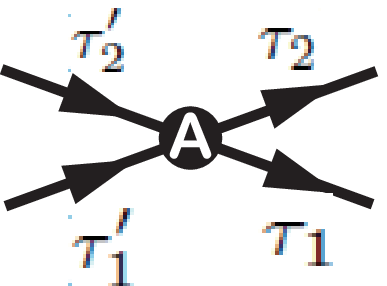}}, \;\;\;\; C^{(0)}_{2B}(\tau^{\prime}_{1},\tau^{\prime}_{2}\mid
\tau_{1},\tau_{2})=\raisebox{-0.7cm}{\includegraphics[width=1.8cm]{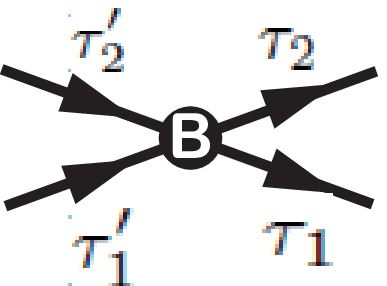}},
\end{equation}
where the pointing-in arrows represent creation operators and the pointing-out arrows represent annihilation operators.

In principle, the one-particle Green's function for the inhomogeneous lattice Bose system may be calculated accurately by counting contribution of all connected diagrams consisting of all nonvanishing cumulants and hoppings. Formally it can be expressed diagrammatically as follows
\begin{equation}\label{inhomogeneous-loop}
G^{\rm inh}(\tau',i|\tau,j)=\raisebox{-0.46cm}{\includegraphics[width=2cm]{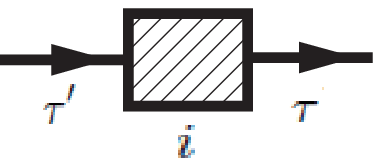}}+
\raisebox{-0.47cm}{\includegraphics[width=3.2cm]{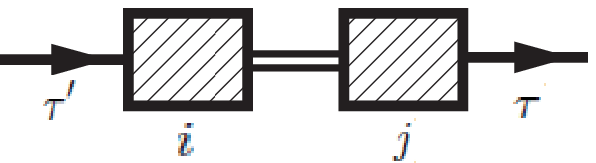}}
+\raisebox{-0.47cm}{\includegraphics[width=4.4cm]{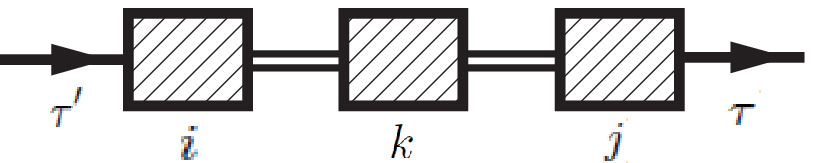}}+\cdots,
\end{equation}
the squares in the diagrams represent the contribution from all non-vanishing cumulants together with hoppings, double-lines  represent all possible hopping connections between nearest-neighbor sites.

In practice, one needs to calculate the Green's function perturbatively by selecting a specific group of diagrams in a proper way. As is well known \cite{Kleinert,Quantum-Field-Theory}, Green's functions should be diverging in the vicinity of second-order phase transition point, however, this property cannot be exposed just by counting finite orders of the perturbation parameter $J/U$, since a polynomial will always be finite, meanwhile, $J/U$ is no longer a small quantity when approaching the phase boundaries. Hence, we re-group the diagrams based on the properties of loops in each diagram rather than the order of $J/U$, each group now consists of infinite number of diagrams, this is the so-called re-summed Green's function technique \cite{ohliger,jiang-pra-2}. At present, for simplicity, we only consider the resummed Green's function $\tilde{G}^{\rm inh}_1$ containing only diagrams of chains consisting of first order cumulants and hopping between sites. In terms of Matsubara frequency, the re-summed Green's function reads
\begin{eqnarray}\label{inhomogeneous-chain}
\tilde{G}^{\rm inh}_1(\omega_{m},i,j)&=&\raisebox{-0.34cm}{\includegraphics[width=1.9cm]{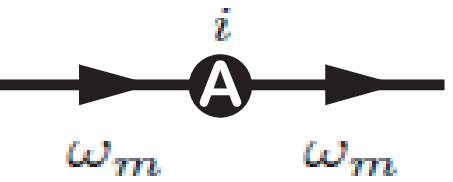}}+
\raisebox{-0.34cm}{\includegraphics[width=2.7cm]{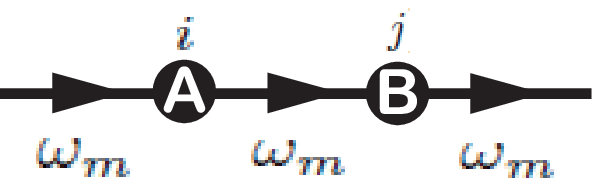}}
+\raisebox{-0.33cm}{\includegraphics[width=3.4cm]{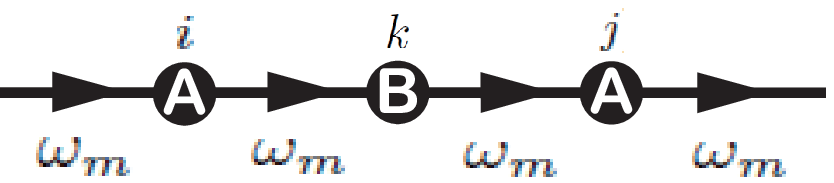}}+\cdots,\nonumber\\
&&+\raisebox{-0.34cm}{\includegraphics[width=1.9cm]{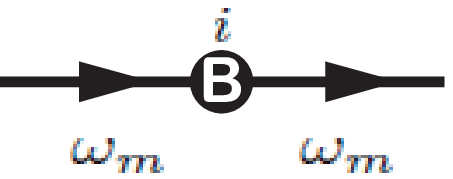}}+
\raisebox{-0.34cm}{\includegraphics[width=2.7cm]{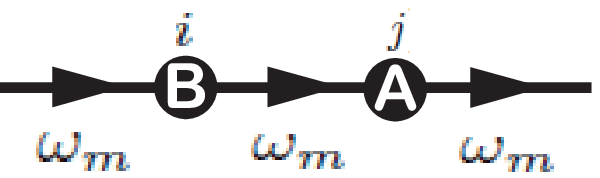}}
+\raisebox{-0.33cm}{\includegraphics[width=3.4cm]{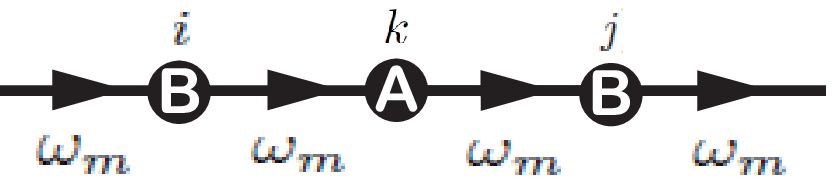}}+\cdots.
\end{eqnarray}
In fact, the present choice of the re-summed Green's function may be looked upon as a sort of mean field treatment \cite{ohliger,jiang-pra-2} and is going to be accurate when the dimension of the system $d$ goes to infinity. Without doubt, from the discussion, it is clear to see that the re-summed Green's function method can in principle go beyond mean field by adding loop diagrams constituted by higher order cumulants and related hoppings.

The above re-summed Green's function diagrams can be re-expressed mathematically in terms of first-order cumulants and hopping parameters in momentum-space as
\begin{eqnarray}\label{inhomogeneous-chain1}
\tilde{G}^{\rm inh}_1(\omega_{m},\textbf{k})&=&C^{(0)}_{1A}(\omega_m)+C^{(0)}_{1A}(\omega_m)C^{(0)}_{1B}(\omega_m)J(\textbf{k})+C^{(0)}_{1A}(\omega_m)^{2}C^{(0)}_{1B}(\omega_m) J^{2}(\textbf{k})+\cdots\nonumber\\
&&+C^{(0)}_{1B}(\omega_m)+C^{(0)}_{1B}(\omega_m)C^{(0)}_{1A}(\omega_m)J(\textbf{k})+C^{(0)}_{1B}(\omega_m)^{2}C^{(0)}_{1A}(\omega_m)J^{2}(\textbf{k})+\cdots\nonumber\\
&=&\frac{C^{(0)}_{1A}(\omega_m)+C^{(0)}_{1B}(\omega_m)+2C^{(0)}_{1A}(\omega_m) C^{(0)}_{1B}(\omega_m) J(\textbf{k})}{1-C^{(0)}_{1A}(\omega_m) C^{(0)}_{1B}(\omega_m) J^{2}(\textbf{k})},
\end{eqnarray}
where
\begin{equation}
C^{(0)}_{1A(B)}(\omega_{m})=\int^{\beta}_{0}C^{(0)}_{1A(B)}(\tau)e^{i\omega_{m}\tau} d\tau
\end{equation}
and
\begin{equation}
J(\mathbf k )= \sum _{ij} J_{ij} e^{i \mathbf k\cdot\mathbf
r_{i}}e^{- i\mathbf k\cdot\mathbf r_{j}}.
\end{equation}
Since $C^{(0)}_{1} (\tau)=\,\langle
\hat{T}_{\tau}[\hat{a}^{\dagger}(\tau) \hat{a}(0)] \rangle_{0}$,
by counting the detailed information of the eigenstates of $\hat{H}_0$, a complicated yet straightforward calculation leads to
\begin{equation}
C^{(0)}_{1A}(\omega_{m})=\frac{1}{Z^{(0)}}\sum^{\infty}_{n_{A},n_{B}=0}\left[\frac{n_{A}+1}{E_{n_{A}+1,n_{B}}-E_{n_{A},n_{B}}+i\omega_{m}}
-\frac{n_{A}}{E_{n_{A},n_{B}}-E_{n_{A}-1,n_{B}}+i\omega_{m}}
\right]e^{-\beta E_{n_{A},n_{B}}},
\end{equation}
and
\begin{equation}
C^{(0)}_{1B}(\omega_{m})=\frac{1}{Z^{(0)}}\sum^{\infty}_{n_{A},n_{B}=0}\left[\frac{n_{B}+1}{E_{n_{A},n_{B}+1}-E_{n_{A},n_{B}}+i\omega_{m}}
-\frac{n_{B}}{E_{n_{A},n_{B}}-E_{n_{A},n_{B}-1}+i\omega_{m}}
\right]e^{-\beta E_{n_{A},n_{B}}},
\end{equation}
where $Z^{(0)}=\sum^{\infty}_{n_{A},n_{B}=0}e^{-\beta E_{n_{A},n_{B}}}$.

Divergence of the Green's function requires
\begin{equation}
1-C^{(0)}_{1A}(\omega_{m}) C^{(0)}_{1B}(\omega_{m})  J^{2}(\textbf{k})=0.
\end{equation}
By taking that the phase transitions occur due to long-wavelength fluctuations \cite{Kleinert,Quantum-Field-Theory} into account, the above formula determines the locations of the phase boundaries of localized states via the equation below
\begin{eqnarray}
z^{2}J_c^{2}&=&\frac{Z^{(0)}}{ \sum^{\infty}_{n_{A},n_{B}=0}\left[\frac{n_{A}+1}{E_{n_{A}+1,n_{B}}-E_{n_{A},n_{B}}}
-\frac{n_{A}}{E_{n_{A},n_{B}}-E_{n_{A}-1,n_{B}}}\right]e^{-\beta E_{n_{A},n_{B}}}}\times \nonumber\\
&&\frac{Z^{(0)}}{ \sum^{\infty}_{n_{A},n_{B}=0}\left[\frac{n_{B}+1}{E_{n_{A},n_{B}+1}-E_{n_{A},n_{B}}}
-\frac{n_{B}}{E_{n_{A},n_{B}}-E_{n_{A},n_{B}-1}}\right]e^{-\beta E_{n_{A},n_{B}}}}.\label{equation20}
\end{eqnarray}

\begin{figure}[h!]
\centering \subfigure[]{{\includegraphics[width=7cm]{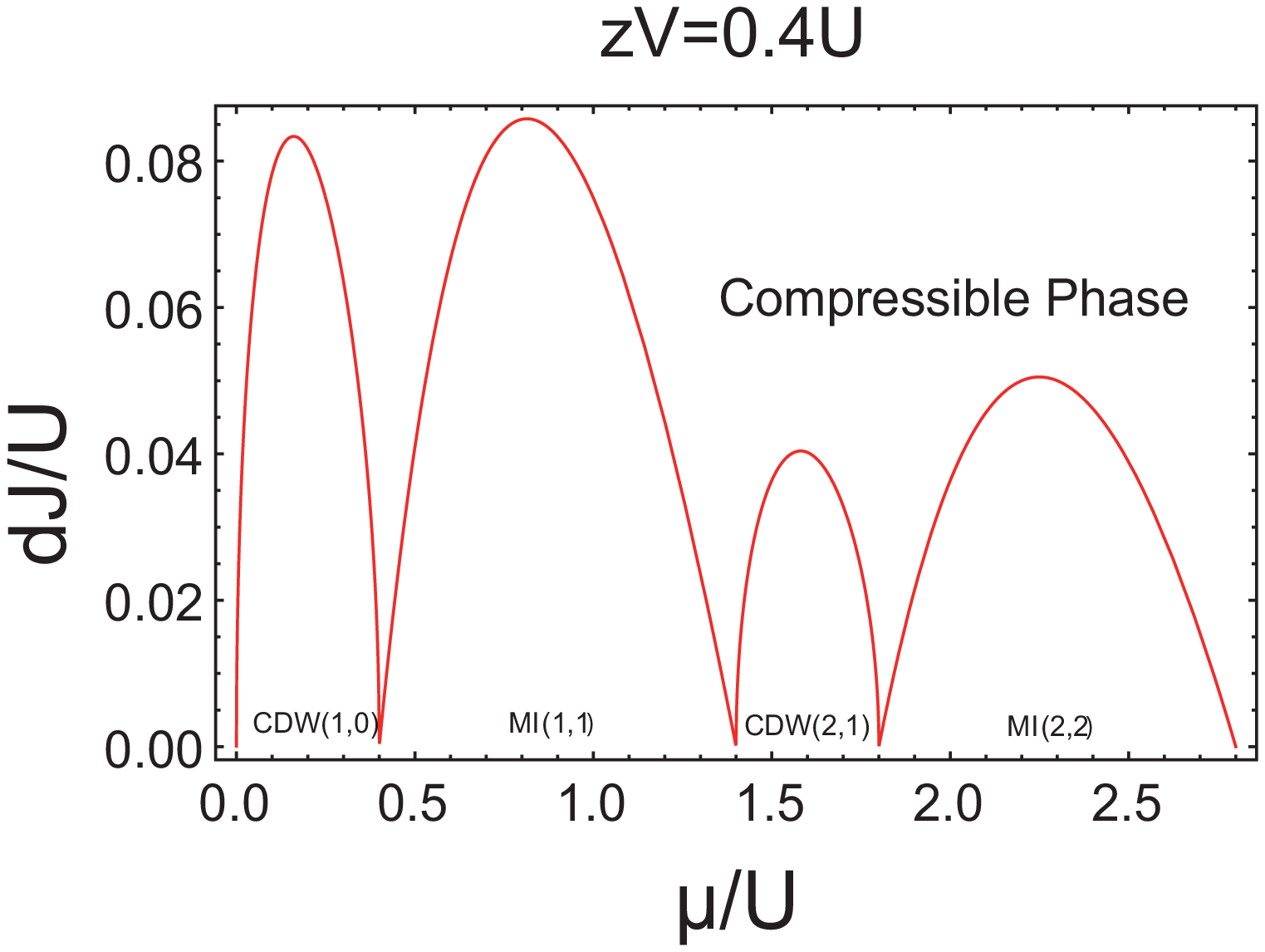}}}
\hspace{0.25in}
\subfigure[]{{\includegraphics[width=7cm]{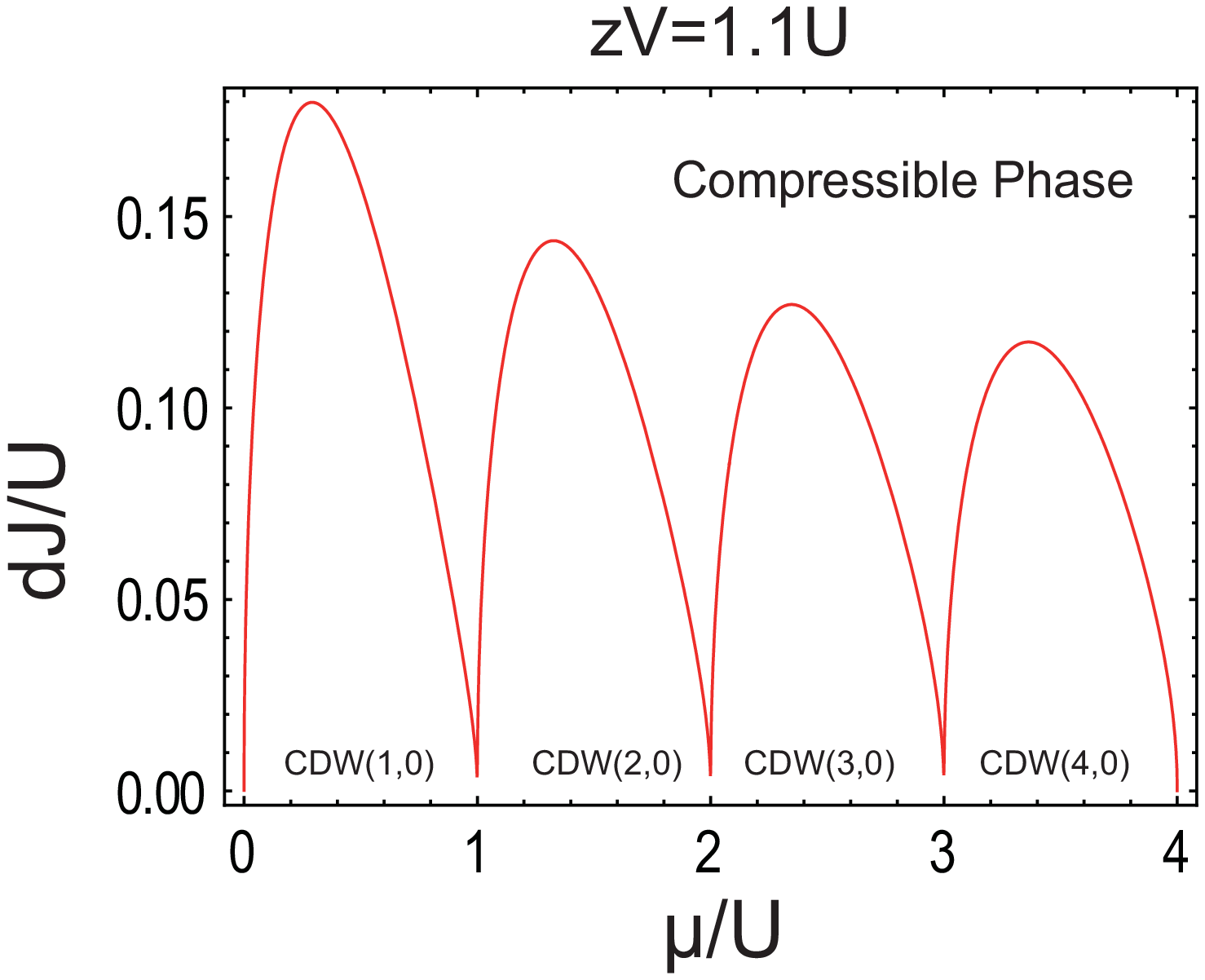}}}
\caption{(Color
online) The quantum phase diagrams of a Bose system in a $d$-dimensional hypercubic lattice with nearest-neighboring repulsive interaction $V$ for cases of (a) $zV=0.4U$ and  (b) $zV=1.1U$.}
\label{phase-diagrams}
\end{figure}

At zero temperature, the eigenstates of the unperturbed part $\hat{H}_0$ fall into the ground states discussed in the preceding section. Suppose that the filling factor of sublattices A and B in the ground state are $n_A$ and $n_B$, the phase-boundary equation Eq.(\ref{equation20}) is then reduced to
\begin{equation}\label{zero-equation1}
z^{2}J_c^{2}=\frac{1}{\left[\frac{n_{A}+1}{E_{n_{A}+1,n_{B}}-E_{n_{A},n_{B}}}
-\frac{n_{A}}{E_{n_{A},n_{B}}-E_{n_{A}-1,n_{B}}}\right]\left[\frac{n_{B}+1}{E_{n_{A},n_{B}+1}-E_{n_{A},n_{B}}}
-\frac{n_{B}}{E_{n_{A},n_{B}}-E_{n_{A},n_{B}-1}}\right]}.
\end{equation}
By substituting the eigenvalues in Eq.(\ref{eigenvalue}) into the above equation, we finally get
\begin{equation}\label{zero-equation2}
\frac{1}{z^{2}J_c^{2}}=\left[\frac{n_{A}+1}{Un_{A}+Z V n_{B}-\mu}
-\frac{n_{A}}{U(n_{A}-1)+Z V n_{B}-\mu}\right]\left[\frac{n_{B}+1}{Un_{B}+Z V n_{A}-\mu}
-\frac{n_{B}}{U(n_{B}-1)+Z V n_{A}-\mu}\right].
\end{equation}
As examples, we show the phase diagrams for the system with $zV=0.4U$ and $zV=1.1U$ in Figs.(\ref{phase-diagrams}a) and (\ref{phase-diagrams}b), respectively.
Our analytical results from extended re-summed Green's function method become exactly the same as the mean field results when $d\rightarrow \infty$ \cite{RPA,Strong-coupling}, consistent with preceding discussion.

\section{Summary}

In summary, we have developed a generalized re-summed Green's function method in this paper. This generalization makes the re-summed Green's function not only a powerful tool to tackle the problems in ultracold Bose systems with homogeneous lattice structures, but also suitable for investigating the quantum critical phenomena in ultracold Bose systems with bipartite sublattice structures. By treating the extended Bose-Hubbard model in a hypercubic lattice with nearest-neighbor interaction as a toy model, we have illustrated the generalized Green's function method in a detail way. Due to the nearest-neighbor repulsive interaction, the hypercubic lattice becomes inhomogeneous and is divided into two sublattices, and the inhomogeneity of the lattice structure invalidates the Green's function method in old fashion. In order to resolve this problem, we have generalized the Green's function method and have treated the cumulants on different sublattice separately during the perturbative cumulant expansion of the Green's function. The present generalization allows us to investigate the quantum phase transitions in the system between incompressible localized states (Mott insulating states or charge density wave states) and compressible delocalized phases (superfluid phase or supersolid phase). To the lowest order, we have calculated out the phase diagrams of the system for different nearest-neighbor interaction $V$. Actually, as discussed before, the lowest order re-summed Green's function is in a sense a sort of mean field theory treatment, however, from Eq.(\ref{inhomogeneous-loop}) we clearly see that the generalized re-summed Green's function method may in principle go beyond mean field by adding loop diagrams consisting of higher order cumulants. Moreover, according to Eq.(\ref{equation20}), to investigate the finite-temperature properties of the system is rather straightforward.

It should be pointed out that the generalized re-summed Green's function method is not only suitable for tackling the quantum phase transitions in hypercubic optical lattice Bose systems with nearest-neighbor interactions, but more importantly, it shed a light on investigating the quantum properties of a Bose system in a bipartite non-Bravais optical lattices like honeycomb lattice, for instance calculating analytically the time-of-flight absorption pictures in this system, which so far has not yet been touched.

\section*{Acknowledgement}

YJ acknowledges Axel Pelster for his stimulating and fruitful
discussions. This Work was supported by National Natural Science Foundation of China [Grant Nos. 11275119 (YJ), 11074043(ZL \& YC) and 11274069(ZL \& YC)] and by Ph.D. Programs Foundation of Ministry of Education of China under Grant No. 20123108110004 (YJ). This work was also supported by the State Key Programs of China (Grant Nos. 2012CB921604 and 2009CB929204) (ZL \& YC).

\end{document}